\documentstyle[prl,multicol,aps]{revtex} \input epsf

\begin{document}

\draft 


\title{Liquid Limits: The Glass Transition and Liquid-Gas Spinodal
Boundaries of Metastable Liquids}

\author{Srikanth Sastry$^{*}$}
\address{Jawaharlal Nehru Centre for Advanced Scientific Research, 
Jakkur Campus, Bangalore 560064, INDIA}                            

\maketitle

\begin{abstract}
The liquid-gas spinodal and the glass transition define ultimate
boundaries beyond which substances cannot exist as (stable or
metastable) liquids. The relation between these limits is analyzed
{\it via} computer simulations of a model liquid. The results obtained
indicate that the liquid - gas spinodal and the glass transition lines 
intersect at a finite temperature, implying a glass - gas mechanical
instability locus at low temperatures. The glass transition lines 
obtained by thermodynamic and dynamic criteria agree very well with
each other.
\end{abstract} 



\begin{multicols}{2}

Substances can exist in the liquid state beyond equilibrium phase
boundaries in a metastable state, if nucleation of the stable phase is
avoided. Such metastable liquids are nevertheless bounded by ultimate
limits in the form of the liquid-gas spinodal and the glass transition
locus. The liquid-gas spinodal is a limit of stability, beyond which
the system is mechanically unstable. Although the nature of the glass
transition, in particular the existence of an ideal glass transition,
is still a matter of debate\cite{reviews}, the locus of glass
transition temperatures defines a boundary beyond which substances
transform to an amorphous solid state, and can no longer exist as
liquids. A preliminary report of investigations concerning the
relationship between these two limits of the liquid state are
presented in this letter.

\begin{figure}[b]
\hbox to\hsize{\epsfxsize=1.0\hsize\hfil\epsfbox{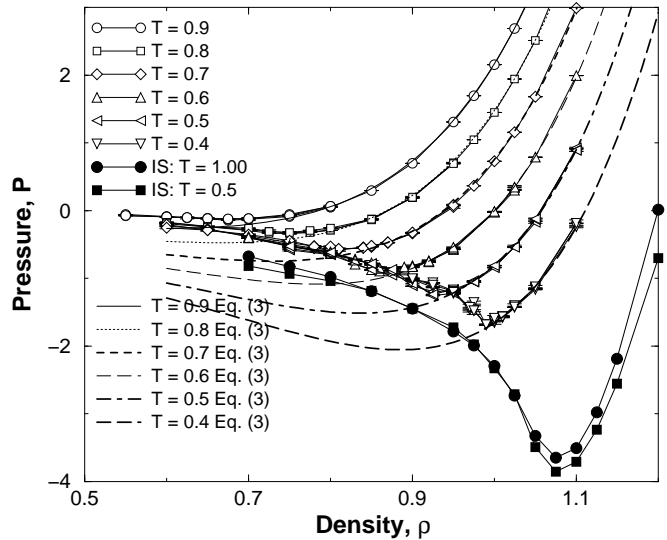}\hfil}
\caption{\narrowtext (a) Isotherms from REMC simulations (points
connected with lines). (b) Curves marked `IS:' are pressure {\it vs.}
density curves for inherent structures, with starting temperatures $T
= 1.0, 0.5$. (c) Continuous curves labeled `Eq. (3)' are from the
empirical equation of state described in the text. }
\label{fig1}
\end{figure}
The immediate motivation for the present study are observations made
in Ref. \cite{sds97}, where a threshold density $\rho^* = 0.89$ was
identified for a model (mono) atomic liquid, across which the structure of
typical local potential energy minima or `inherent structures' (IS)
undergoes a qualitative change from spatially heterogeneous (below
$\rho^*$) to homogeneous structures (above $\rho^*$). At the same
density $\rho^*$, the pressure $P$ {\it vs.} density curve for the
inherent structures goes through a minimum. Two speculations were made
in \cite{sds97}, motivated by these observations: (1) The threshold
density $\rho^*$ is (or closely approximates) the $T \rightarrow 0$
limit of the liquid-gas spinodal locus $\rho_s(T)$, and (2) The
threshold density $\rho^*$ forms the absolute lower density limit to
glass formation. In the simplest scenario, the glass transition
temperature approaches zero as $\rho \rightarrow \rho^*$ from
above. These speculations are tested here {\it via} computer
simulations of a model liquid.  The results obtained also prove
valuable in assessing recent approaches to studying the glass
transition.

The model studied is a binary mixture of $204$ type $A$ and $52$ type
$B$ particles, interacting {\it via} the Lennard-Jones (LJ) potential,
with parameters $\epsilon_{AB}/\epsilon_{AA} = 1.5$,
$\epsilon_{BB}/\epsilon_{AA} = 0.5$, $\sigma_{AB}/\sigma_{AA} = 0.8$,
and $\sigma_{BB}/\sigma_{AA} = 0.88$, and $m_B/m_A = 1$. This system
has been extensively studied as a model glass
former\cite{kob,sastry,fs,parisi}. Monte Carlo (MC) simulations in the
restricted ensemble (REMC) were performed (details below) for $8$
temperatures for an average of $10$ densities, for three different
sets of constraints in each case, to locate the liquid-gas spinodal
locus. Run lengths ranged from $3 \times 10^5$ to $1.8 \times 10^6$ MC
cycles. Molecular Dynamics (MD) simulations (details as in
\cite{sastry}) were performed at $7$ densities, $\rho = 1.08, 1.1,
1.125, 1.15, 1.2, 1.25, 1.35$, over a wide range of temperatures (from
$T = 3.0$ to $0.259$ for $\rho = 1.08$). Run lengths ranged from $2
\times 10^5$ for $T > 1.0$ to $6.2 \times 10^7$ time steps
(or $1.86 \times 10^5$ LJ time units, or $0.4~\mu s$ in Argon
units). Simulations above $T = 1.0$ were typically done at constant
temperature (NVT) and those below $T = 1.0$ at constant energy
(NVE). NVT simulations were also performed for lower densities ($\rho
= 0.55$ to $1.05$, for $T = 1.0$ to $0.325$) to obtain isothermal
compressibilities, $k_T$.

The liquid-gas spinodal is estimated {\it via} restricted ensemble
simulations\cite{resens,corti}, wherein the system is divided into
cells and fluctuations in density in each cell are
restricted. Isotherms so obtained display a van der Waals-type loop,
and permit estimation of the spinodal density from the location of
their minima (See \cite{corti} for further details)\cite{fn1}. The
resulting isotherms are displayed in Fig. 1, along with the $P$ {\it
vs.}  $\rho$ curves for inherent structures, obtained at two
temperatures. The latter show that IS properties depend on the
starting $T$, but the density at the minimum is not significantly
altered, and is $\rho^* = 1.08$. Finite $T$ isotherms do not show
significant dependence on constraint strength. Spinodal densities
$\rho_s(T)$ obtained from the location of minima along these
isotherms, for each constraint strength, are shown (as spinodal
temperatures $T_s(rho)$) in Fig. 2. Also shown is $\rho^*$ where the
IS pressure is minimum. To confirm the reliability of the REMC
estimates, the spinodal is also estimated (a) by fitting isotherms
($P$ {\it vs.} $\rho$) by (cubic) polynomials, and (b) by calculating
$k_T$ directly in MD simulations, and locating the density where
$k_T^{-1}$ vanishes by polynomial extrapolation. The results are
shown in Fig. 2, which shows that spinodal densities so obtained
agree well with the REMC estimates.
\begin{figure}
\hbox to\hsize{\epsfxsize=1.0\hsize\hfil\epsfbox{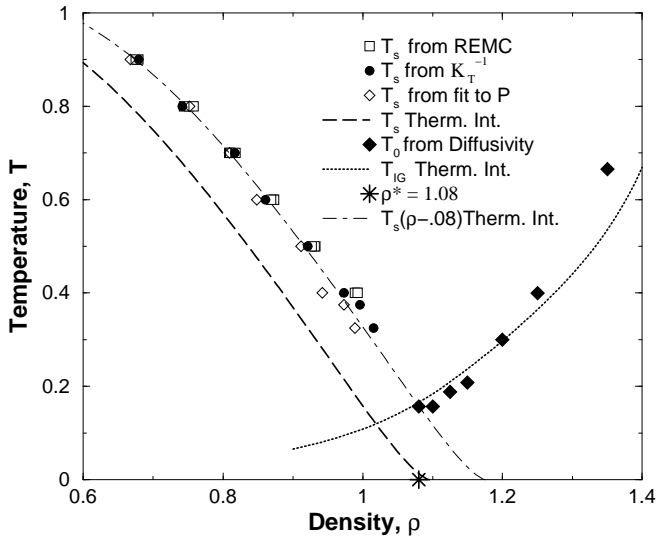}\hfil}
\caption{\narrowtext Liquid-gas spinodal obtained from (a) REMC
simulations, (b) $k_T$, (c) polynomial fits to isotherms, and (d) the
empirical free energy (`$T_s$ Therm. Int.'). The same curve is
also shown shifted in $\rho$ by 0.08 (`$T_s (\rho-.08)$
Therm. Int.'). The glass transition locus obtained from (e) VFT fits
to diffusivity data, and (d) extrapolation of configurational entropy
to zero (`$T_{IG}$ Therm. Int.'). Also marked (*) is the
density $\rho^*$ where inherent structure pressure is a minimum.}
\label{fig2}
\end{figure}
An empirical free energy is constructed next, based on equilibrium
data from MD simulations\cite{fs,parisi,speedy,speedyag,scala}, which
is used both to obtain an independent estimate of the spinodal, and to
obtain a thermodynamic estimate of the glass transition locus.  The
absolute free energy $A(\rho, T)$ of the system at density $\rho$ at a
reference temperature $T_r = 3.0$ is first defined in terms of
the ideal gas contribution $A_{id}(\rho, T)$ and the excess free
energy $A_{ex}(\rho, T)$ obtained by integrating the pressure from
simulations:
\begin{eqnarray} 
A(\rho,T) = A_{id}(\rho,T) + A_{ex}(\rho,T),\\
\beta A_{id}(\rho,T)  = N \left(3~\ln \Lambda + \ln \rho - 1\right),\nonumber \\
\beta_r A_{ex}(\rho,T_r) = \beta_r A^{0}_{ex}(0,T_r) + N  
\int_{0}^{\rho} {d\rho^{'} \over \rho^{'}}\left( { \beta_r P\over \rho^{'} } - 1\right),\nonumber \\
\beta_r A^{0}_{ex}(0,T_r) = - \ln {N! \over N_A! N_B!}.\nonumber
\end{eqnarray}
Here, $N$ is the number of particles, $\beta \equiv k_B T$, $\Lambda$
is the de Broglie wavelength, and $A^0_{ex}$ arises from the mixing
entropy. $A_{ex}$ is fit to a fifth order polynomial is
$\rho$\cite{fits}. $A_{ex}$ at a desired temperature may
be evaluated by integrating the potential energy, $E$:
\begin{equation} 
\beta A_{ex}(\rho, \beta) =  \beta A_{ex}(\rho,\beta_r) + 
\int_{\beta_r}^{\beta} E(\rho, \beta^{'}) d\beta^{'} 
\end{equation}
As observed in \cite{fs,parisi}, the $T$ dependence of $E$
at fixed, high, density is well described by the form
$E(\rho, T) \sim T^{3/5}$, in agreement with predictions for dense
liquids\cite{rosetar}. This form is not, however, accurate at low
densities. In order to fit data well at low densities while retaining 
reliability at high densities, $E$ data for $\rho
= 0.55$ to $1.35$, and temperatures below $T = 3.0$ are fitted to the
form $E(\rho, T)/N = E_0(\rho) + E_1(\rho) T^{E_2(\rho)}$. The
parameters $E_0, E_1, E_2$ are fitted polynomials in
$\rho$\cite{fits}. These fits, which represent the measured MD data
very well, together with the fit for $A^{0}_{ex}(0,T_r = 3.0)$, define
an empirical free energy from which the equation of state and the
liquid-gas spinodal locus $\rho_s(T)$ are obtained {\it via}:
\begin{eqnarray}
P(\rho,T) = {\rho^2 \over N} {\partial  A(\rho,T)\over \partial \rho}~ \vline_{_{_{_{~ T}}}}; ~~~ {\partial P \over \partial \rho} ~\vline_{_{_{~ \rho = \rho_s(T)}}} = 0. 
\label{eos}
\end{eqnarray}

The liquid-gas spinodal locus resulting from the empirical free energy
is shown in Fig. 2 and is seen to occur at lower densities than the
REMC estimate, with a roughly constant shift in density (also shown in
Fig. 2). Inspection of isotherms obtained from the empirical free
energy (which show mean field behavior near the spinodal) reveals that
while they agree very well with simulation isotherms away from the
spinodal, they show deviations very close to the (REMC) spinodal
densities. Correspondingly $k_T^{-1}$ obtained in simulations
drops faster to zero close to the spinodal, while at higher densities
there is very good agreement.  The estimate from the empirical free
energy must thus be considered a lower limit to the location of the
spinodal.

Glass transition temperatures at the $7$ studied densities are
obtained next, by fitting diffusion coefficients of $A$ particles to
the Vogel-Fulcher-Tammann-Hesse (VFT) form:
\begin{equation}
D(\rho,T) = D_o(\rho) exp\left({A(\rho)\over T - T_0(\rho)}\right).
\end{equation}
Values of $T_0(\rho)$ so estimated define lower limits to the
laboratory glass transition $T_g$. The diffusion coefficients and the
corresponding VFT fits are shown in Fig. 3. The resulting glass
transition temperatures $T_0(\rho)$ are shown in Fig. 2. (Relaxation
times yield very similar estimates. A qualitatively similar $T_g$ {\it
vs.} $\rho$ curve has recently been reported in \cite{ruocco2}.)

\begin{figure}
\hbox to\hsize{\epsfxsize=1.0\hsize\hfil\epsfbox{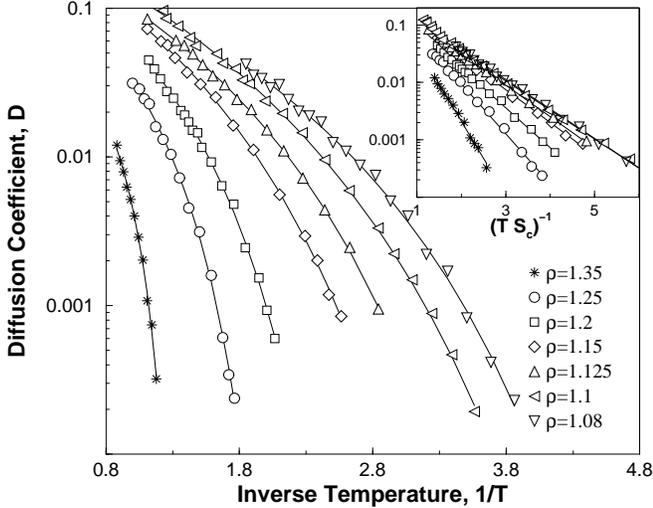}\hfil}
\caption{Arrhenius plot of diffusion coefficients and the corresponding VFT fits for the $7$ densities studied. (inset) Adam-Gibbs plot of diffusion coefficients against $(T S_c)^{-1}$ and straight line fits.}
\label{fig3}
\end{figure}
A thermodynamic estimate of ideal glass transition temperatures is
obtained next, by evaluating the configurational entropy of the liquid
as a function of density and temperature. A number of studies in
recent
years\cite{fs,parisi,speedy,speedyag,scala,ruocco,silvio,mezard,parisi2,heuer,jpcpgd}
have focussed on evaluating of the configurational
entropy\cite{adam-gibbs}. The approach here, based on analyzing
inherent structures\cite{inh}, follows most closely those of Ref.s
\cite{fs,heuer,parisi,parisi2}. The canonical partition function is re-written
as a sum over all local potential energy minima, which introduces a
distribution function for the number of minima at a given energy:
\begin{eqnarray}
Q_N(\rho,T) = \Lambda^{-3N} {1\over N_A! N_B!} \int d{\bf r}^N exp\left(-\beta \Phi\right)\\
	    = \sum_\alpha exp\left(-\beta \Phi_\alpha\right) \Lambda^{-3N} \int_{V_\alpha} d{\bf r}^N exp\left(-\beta(\Phi-\Phi_\alpha)\right)\nonumber \\
	    = \int d\Phi_\alpha ~ \Omega(\Phi_\alpha) ~exp\left(-\beta(\Phi_\alpha + N f_{basin}(\Phi_\alpha,T))\right)\nonumber \\
	    = \int d\Phi_\alpha ~exp\left(-\beta (\Phi_\alpha + N f_{basin}(\Phi_\alpha,T) - T S_c(\Phi_\alpha))\right)\nonumber
\end{eqnarray}
where $\Phi$ is the total potential energy of the system, $\alpha$
indexes individual inherent structures, $\Phi_\alpha$ is the potential
energy at the minimum, $\Omega(\Phi_\alpha)$ is the number density of inherent
structures with energy $\Phi_\alpha$, and the configurational entropy $S_c
\equiv k_B \ln \Omega$. The {\it basin free energy}
$f_{basin}(\Phi_\alpha,T)$ is obtained by a restricted partition function sum
over a given inherent structure basin, $V_\alpha$. In the harmonic approximation, we have
\begin{equation}
\beta f  = {3\over 2} \ln({\beta \over2\pi}) + {1 \over 2N} \sum_i^{3N-3} \ln \lambda_i \equiv \beta f_{therm} + \beta f_{vib},
\end{equation}
where $\lambda_i$ are eigenvalues of the Hessian or curvature matrix
at the minimum. $\beta f_{vib}$ is a slowly varying function of
temperature (the temperature dependence is obtained by averaging over
$1000$, $100$ inherent structures at $T < 1.$, $T>1.$ respectively),
and is fitted to the form $\beta f_{vib}(\rho,T) = f_0(\rho) +
f_1(\rho)/T^2$ which fits available data quite well. Polynomial fits
to $\rho$-dependent parameters $f_0, f_1$ are obtained\cite{fits}. The
total entropy of the liquid $S$ as well as the basin entropy
$S_{basin}$ may be evaluated as a function of density and temperature
from the total and basin free energies. The configurational entropy
$S_c(\rho, T)$ and the ideal glass transition $T_{IG}(\rho)$ are then
given by,
\begin{eqnarray}
S_c(\rho, T) = S(\rho,T) - S_{basin} (\rho,T);S_c(\rho,T_{IG}(\rho)) = 0.
\end{eqnarray}
The ideal glass transition locus obtained is shown in Fig. 2, and is
seen to be in very good agreement with the estimate based on
diffusivity. Although such correlation is recorded for experimental
data\cite{tkto}, it is noteworthy that $T_0$ estimates here from
dynamic data at quite high $T$ produce such agreement. This agreement
offers mutual support to the thermodynamic method followed here, and
the feasibility of estimation based on dynamical
measurements. $T_{IG}(\rho =1.2) = 0.2976$ is in very good agreement
with the values $0.297$ in \cite{fs} and $0.31$ in \cite{parisi}.
Diffusion coefficients plotted against $(T S_c)^{-1}$ (inset of
Fig. 3) show that the Adam-Gibbs expectation\cite{adam-gibbs} $\ln{D}
\sim (T S_c)^{-1}$ is extremely well satisfied at all
densities\cite{speedyag,scala,tkto}.

Data in Fig. 2 clearly indicate that the liquid-gas spinodal and glass
transition loci would intersect at a finite temperature $T_i \sim
0.16$ (based on REMC data; $T_i \sim 0.12$ based on the empirical free
energy). As this intersection happens at $\rho \sim \rho^* = 1.08$,
the data do not support the speculation that as $T \rightarrow 0$,
$\rho_s(T) \rightarrow \rho^*$ or $\rho_{IG}(T) \rightarrow
\rho^*$. On the other hand, $\rho^*$ does (within the uncertainty of
the data) form the lower limiting density for glass
formation. Experimental data (see {\it e. g.}  \cite{tgvsp}) for the
P-dependence of $T_g$ indicate that $T_g(P)$ typically intersects the
zero pressure axis, implying glass - gas coexistence and finite $T_g$
at negative pressures, a possibility implicit in \cite{sds97} and also
noted in \cite{parisi2}. However, intersection of the liquid-gas
spinodal and the glass transition locus at a finite temperature
observed here further implies an ideal glass - gas mechanical
instability below $T_i$. Indirect evidence for this exists in the form
of experimental $T_g(P)$ loci displaying a negative slope\cite{tgvsp},
but systematic evidence requires vitrification experiments at
negative pressures\cite{tgnegp}.

A proper study of such a mechanical instability is beyond the scope of
the present work, as it requires calculation of limits of stability
accounting consistently for the presence of the three relevant phases
(liquid, gas, glass). As a preliminary attempt, the equation of state
of the glass below $T_{IG}(\rho)$ is obtained by noting that below
$T_{IG}$ the system is trapped in the potential energy basin reached
at $T_{IG}$. Details of such a calculation will not be presented here;
Instead, two consequences are presented in Fig. 4.  The inset
demonstrates the slope discontinuity of the pressure at fixed density
at $T_{IG}$ for $\rho=1.2$.  The main diagram shows the isotherm
at $T = 0.05$ for the glass phase, which displays a mechanical
instability density of $\rho \sim 1.$, where the slope vanishes, or,
$k_T$ diverges.
\begin{figure}
\hbox to\hsize{\epsfxsize=1.0\hsize\hfil\epsfbox{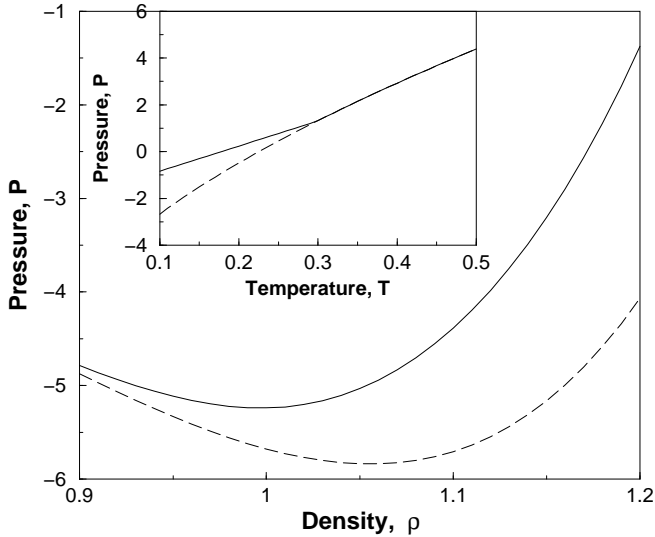}\hfil}
\caption{Pressure {\it vs.} density for $T=0.05$ in the ideal glass
phase (solid line), which exhibits a minimum at $\rho \sim
1.$. Extension of the liquid's equation of state is shown for
comparison (dashed line). Inset shows pressure {\it vs.} temperature
across the ideal glass transition for $\rho = 1.2$ (solid line).}
\label{fig4}
\end{figure}
An interesting by-product of the present analysis is the estimation of
configurational entropy as a function of both energy and density, and
a range of temperature dependences of dynamical quantities displaying
a variable degree of fragility\cite{fragility}. These data permit an
evaluation of the relationship between fragility and configurational
entropy\cite{speedyfr}, and will be presented elsewhere. 

Useful discussions with and comments on the manuscript by
C. A. Angell, D. S. Corti, C. Dasgupta, P. G. Debenedetti, F. Sciortino,
R. J. Speedy and F. H. Stillinger are gratefully acknowledged.

\end{multicols}


\begin{references}
\bibitem{fits} The fit parameters for various quantities are:
$A_{ex}(\rho,T_r=3.0)/(Nk_B T_r) = -0.504704+0.146886 \rho +0.593845
\rho^2-0.080963 \rho^3+0.469425 \rho^4+0.106531 \rho^5$, $E_0(\rho) =
6.52209-38.7893~\rho+60.0495~\rho^2-49.5606~\rho^3+15.084~\rho^4 $,
$E_1(\rho ) = -3.7406+19.6936~\rho-32.6655~\rho^2+
25.0497~\rho^3-6.5492~\rho^4$, and $E_2(\rho) =
0.402798+0.26946~\rho-0.0789979~\rho^2$, 
$f_0(\rho) = 69.2709 -162.481 \rho+137.954 \rho^2-37.4498 \rho^3$,
$f_1(\rho) = 0.581045- 1.76403 \rho+ 1.76942 \rho^2- 0.595709 \rho^3$.
\end{references}
\end{document}